\documentclass[journal=jacsat,manuscript=article]{achemso}
\usepackage{chemformula} 
\usepackage[T1]{fontenc} 

\author{Yulu Ren}
\affiliation{State Key Laboratory of Metastable Materials Science and Technology \& Key Laboratory for Microstructural Material Physics of Hebei Province, School of Science, Yanshan University, Qinhuangdao 066004, China}
\author{Qiaoqiao Li}
\affiliation{State Key Laboratory of Metastable Materials Science and Technology \& Key Laboratory for Microstructural Material Physics of Hebei Province, School of Science, Yanshan University, Qinhuangdao 066004, China}
\author{Wenhui Wan}
\affiliation{State Key Laboratory of Metastable Materials Science and Technology \& Key Laboratory for Microstructural Material Physics of Hebei Province, School of Science, Yanshan University, Qinhuangdao 066004, China}
\author{Yong Liu}
\affiliation{State Key Laboratory of Metastable Materials Science and Technology \& Key Laboratory for Microstructural Material Physics of Hebei Province, School of Science, Yanshan University, Qinhuangdao 066004, China}
\author{Yanfeng Ge}
\email{yfge@ysu.edu.cn}
\affiliation{State Key Laboratory of Metastable Materials Science and Technology \& Key Laboratory for Microstructural Material Physics of Hebei Province, School of Science, Yanshan University, Qinhuangdao 066004, China}

\title[An \textsf{achemso} demo]
  {High-Temperature Ferromagnetic Semiconductors: Janus Monolayer Vanadium Trihalides}

\abbreviations{IR,NMR,UV}

\begin{document}


\begin{abstract}
  Two-dimensional (2D) intrinsic ferromagnetic semiconductors are expected to stand out in the spintronic field. Recently, the monolayer VI$_{3}$ has been experimentally synthesized but the weak ferromagnetism and low Curie temperature ($T_C$) limit its potential application. Here we report that the Janus structure can elevate the $T_C$ to 240 K. And it is discussed that the reason for high $T_C$ in Janus structure originates from the lower virtual exchange gap between $t_{2g}$ and $e_{g}$ states of nearest-neighbor V atoms. Besides, $T_C$ can be further substantially enhanced by tensile strain due to the increasing ferromagnetism driven by rapidly quenched direct exchange interaction. Our work supports a feasible approach to enhance Curie temperature of monolayer VI$_{3}$ and unveils novel stable intrinsic FM semiconductors for realistic applications in spintronics.
\end{abstract}

\section{1. INTRODUCTION}
Two-dimensional materials provide a crucial platform to investigate novel physics and show tremendous application prospects in spintronic and flexible devices due to their greater conductivity and tunability compared with the traditional bulk materials.\cite{AKGeim2013, KSNovoselov2004, QHWang2012, VNicolosi2013, TJWilliams2015} However, two aspects are considered to limit the progress in spintronics. The one is the scanty materials possessing ferromagnetic (FM) order and semiconducting properties. The other one is the fragile nature of 2D magnetic materials.\cite{ChengGong2019, Kyunghee2016, HeTian2016, XDuan2015, Gibertini2019} To solve the problems, one can make the 2D materials precisely tunable by inducing magnetism and phase transition to enlarge the candidates for FM semiconductors. Meanwhile, designing or finding novel and stable magnetic semiconductors are also effective ways.

As an important member of 2D ferromagnetic semiconductors, the monolayer CrI$_3$ with a $T_C$ = 45 K has been reported for their brilliant tunability\cite{Huang2017, WB2015}, such as the layer and gate voltage dependent FM-antiferromagnetism (AFM) transition.\cite{Huang2018, MA2015, Song2019, Jiang2019} Besides, the bilayer Cr$_{2}$Ge$_{2}$Te$_{6}$ has a $T_C$ of 28 K which can be modestly elevated to 44 K by a magnetic field of 0.3 T.\cite{Gong2017} Similarly, in monolayer Fe$_{3}$GeTe$_{2}$, an ionic gate voltage of 1.75 V can drastically raise the $T_C$ from 130 K to room temperature.\cite{DengY2018, FeiZ2018} For applicational interests, it is highly expected that strong ferromagnetism gives rise to stable FM spin array which can survive at the room temperature.\cite{HOhno2010, KAndo2006, PSharma2003, XMarti2014, HJZhao2014} However, $T_C$ in Cr-based honeycomb transition-metal trihalides are limited under $\sim$70 K due to their dilute-magnetism nature, unlike the strong ferromagnetism driven by double-exchange interaction in monolayer Fe$_{2}$Si ($T_C$=$\sim$780 K)\cite{YSun2017, CZener1951}. As a consequence, the great efforts have been put into finding the approaches to boost $T_C$ in Cr-based 2D materials, such as doping, alloying and straining.\cite{SJiang2018, CHuang2018, SChen2019, HLi2019}

Recently, a novel layered van der Waals FM semiconductor VI$_{3}$ has been synthesized via chemical vapor transport method with a $T_C$ = 50 K. It supports a new platform to investigate 2D magnetism and van der Waals heterostructures in $S = 1$ system,\cite{Tian2019, Kong2019, An2019, LiuY2020} but the low $T_C$ is still the considerable disadvantage. In order to find a feasible measure to elevate the $T_C$ of monolayer VI$_{3}$, we take into consideration the structural symmetry breaking, which has initiated a vast amount of interest in modulating the magnetic and electronic properties.\cite{Angyu2017, XPang2014, YHao2019, GDing2019} For example, It has been found $T_C$ of Janus monolayer VSSe reaches as high as 400 K and the stability is comparable with its pristine VSe$_{2}$.\cite{JHe2018, CZhang2019} Also utilizing the strategy, some novel stable FM semiconductors have been predicted such as Janus monolayer X$_{3}$-Cr$_{2}$-Y$_{3}$.\cite{Moaied2018}

In this work, a systematical study is given on Janus monolayers of VI$_{3}$: V$_2$Cl$_3$I$_3$ (VClI), V$_2$Br$_3$I$_3$ (VBrI) and V$_2$Cl$_3$Br$_3$ (VClBr), which are collectively called VXY (X, Y = Cl, Br and I, X $\neq$ Y), based on first-principle calculations. The monolayer VXY is demonstrated that their free-standing films would remain stable experimentally by the stability analysis. Based on Monte Carlo simulations with Heisenberg model, the $T_C$ is estimated drastically surpass monolayer VI$_{3}$ due to the lower virtual exchange gap ($G_{ex}$). Furthermore, a biaxial tensile strain (6\%) can elevate the $T_C$ to 280 K in monolayer VClI. Because when it is stretched, the direct exchange interplay is quenched with the slower weakened superexchange one, leading to FM domination and the elevatory $T_C$.
\section{2. METHODS}
The first-principle calculations were performed using the projected augmented-wave method as implemented in the Vienna Ab initio Simulation Package (VASP).\cite{1, 2} The structural optimizations and convergence tests were adopted with Perdew-Burke-Ernzerhof (PBE) functional,\cite{3} and the band structures were given by Heyd-Scuseria-Ernzerhof (HSE06) functional including 25\% non-local Hartree-Fock exchange. Firstly, $3\times3\times1$ supercell model was applies to perform molecular-dynamic simulations and phonon-frequency calculations. Also, $2\times2\times1$ was used during simulations of different spin arrays and unit cell during the other analysis. Secondly, for some tunable parameters, we kept the following parameters the same from beginning to the end in each analysis: plane-wave cutoff energy, width of smearing, number of k-points in reciprocal space and the thickness of vacuum slab, which were set to 450 eV, 0.05 eV, $9\times9\times1$ and 20 {\AA}, respectively. Note that the parameters were all taken convergence tests against total energy (Figure S1). Thirdly, force and total energy convergence criteria were set to 0.1 meV/{\AA} and $10^{-9}$ eV during relaxations, 20 meV/{\AA} and $10^{-6}$ eV in HSE06 band structures and 1 meV/{\AA}, $10^{-8}$ eV for the other calculations. At last, the vibrational property analyses were obtained with the Phonopy\cite{4} code using the density functional perturbation theory (DFPT)\cite{5} and the molecular dynamic (MD) simulations in the canonical (NVT) ensemble were performed at 300 K with a Nos\'{e} thermostat.
\section{3. RESULTS AND DISCUSSION}
\subsection{3.1 Stability and Feasibility of VXY}
The compounds VXY has layered BiI$_{3}$-type structures\cite{7, 8, 9} and formed by stacks of halogen X-vanadium-halogen Y sandwich layers with lattice parameters a = b = 6.62 {\AA} in monolayer VClI (Section S\uppercase\expandafter{\romannumeral2} in the supplemental material). Figure \ref{1}a,b show that each V atom in a VXY unit cell is nearest-neighbored by three up-layer X and three down-layer Y atoms, forming a V-centered octahedron. We begin with establishing the formation energy ($E_{f}$) to evaluate the stability of VXY. The $E_{f}$ of -0.826 eV in VClI is very close to its pristine VI$_3$ (-0.891 eV), and more stable than monolayer CrGeTe$_{3}$ (-0.588 eV). Then the MD simulation and phonon spectra are employed. Figure \ref{1}c shows VClI can remain free-standing monolayers with honeycomb lattice at room temperature. Apart from the thermal stability from the MD simulation, phonon dispersion relation in Figure \ref{1}d shows the dynamic stability with the absence of imaginary frequency. Nextly, elastic stiffness tensors $C_{11}$ ($C_{11} = C_{22}$), $C_{12}$ and $C_{66}$ are established as 32.56, 12.65 and 9.97 N/m, respectively, indicating the mechanical stability according to Born criterion for hexagonal-structure ($C_{11} > 0$ and $C_{11} - C_{12} > 0$). Young's module and Poisson's rate of monolayer VClI are evaluated with values of 28.2 N/m and 0.39, respectively. The values are smaller than those of some 2D materials, such as graphene (1000 N/m) and MoS$_{2}$ (340 N/m).\cite{11, 12} Thus, the softer monolayer VClI can sustain more strains (see Section S\uppercase\expandafter{\romannumeral2} in the supplemental material for details).
\begin{figure}[htbp]
    \centering
    \includegraphics[scale=0.11]{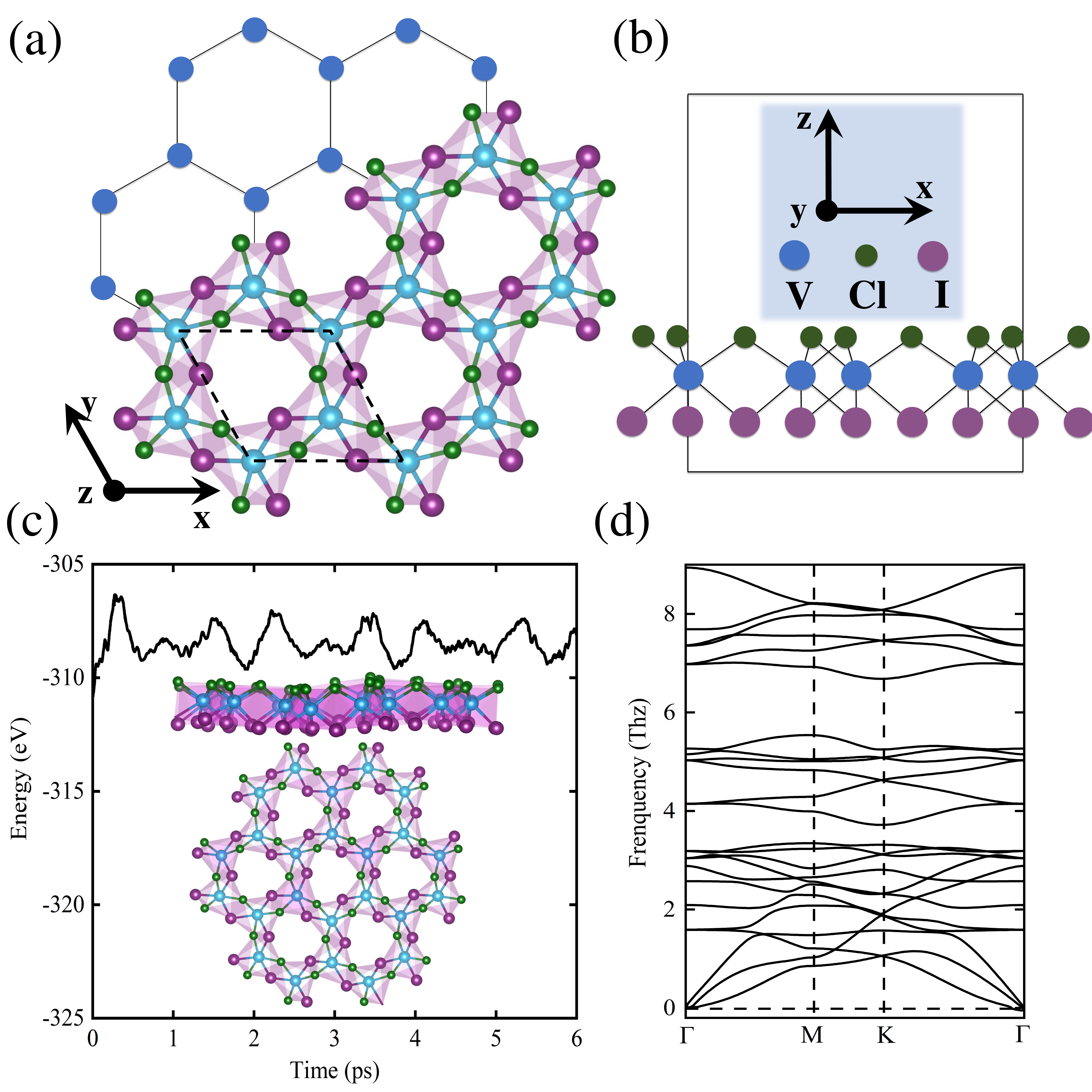}
       \caption{\label{1}(a) Top and (b) side view of hexagonal structural monolayer VClI. The dashed area denotes a unit cell. (c) MD simulation of monolayer VClI at 300 K and the insets are crystal structure after 6 ps. The simulation processes 3000 steps and 2 fs per step. (d) Phonon band structure of monolayer VClI.}
     \end{figure}

\subsection{3.2 Electronic Properties of VXY}
\begin{figure*}[htbp]
    \centering
    \includegraphics[scale=0.15]{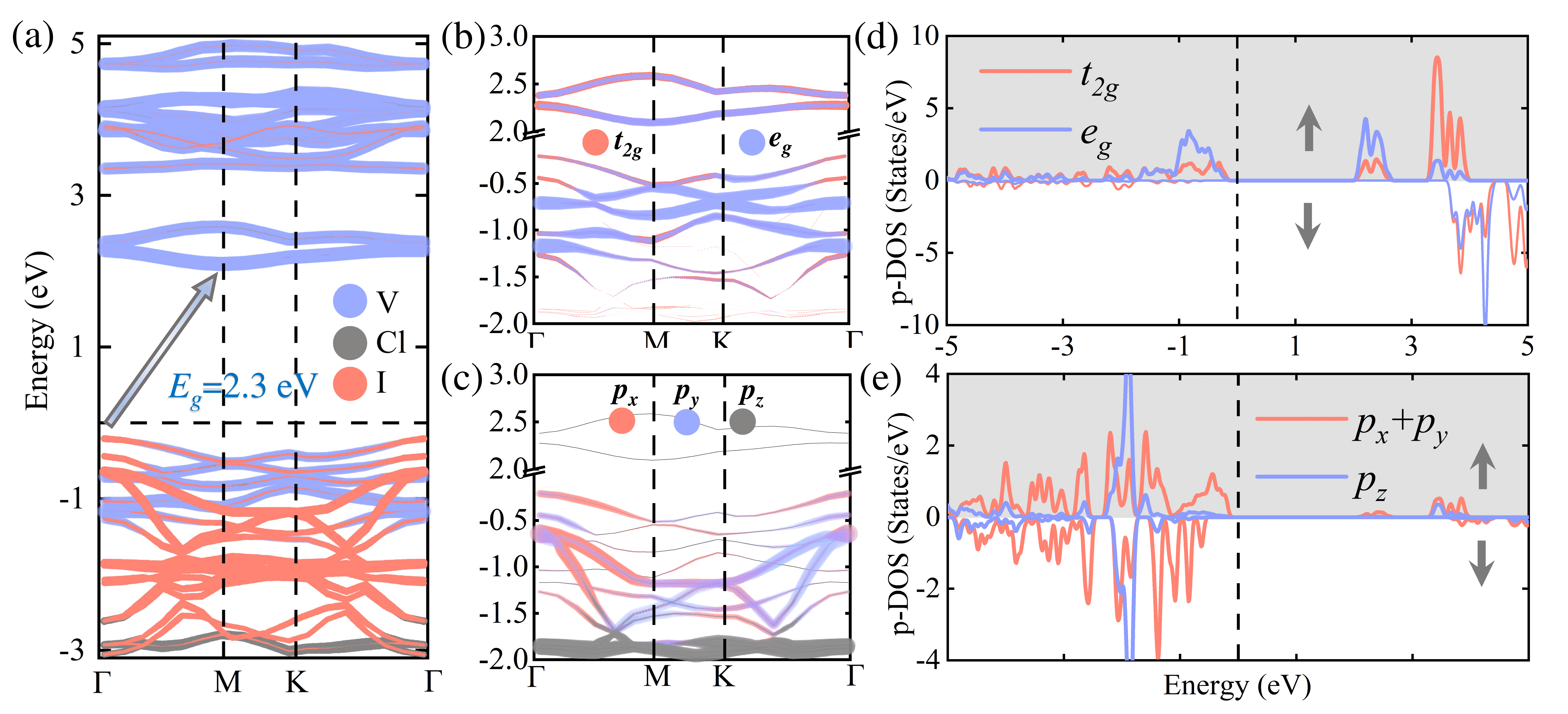}
       \caption{\label{2}(a) Band structure for monolayer VClI. Projected band structures for (b) $d$ orbits of V and (c) $p$ orbits of I atoms. Projected density of states (p-DOS) for (d) $d$ orbits of V and (e) $p$ orbits of I. Gray denotes spin-up areas.}
\end{figure*}

In V-based octahedral environment, $d$ orbits are split into higher degenerate $e_{g}$ states ($d_{x^{2}-y^{2}}$, $d_{z^{2}}$) and lower three-fold-degeneracy $t_{2g}$ states ($d_{xy}, d_{zx}, d_{yz}$). Different from Cr-based trihalides, two electrons are expected to occupy the $t_{2g}$ states and two of three degenerate $t_{2g}$ are half filled, leading to $S = 1$ and valence electronic configuration of $3d^{2}$ with a magnetic moment of 2 $\mu_{B}$ per V atom.\cite{McGuire2017, SSon2019} The band structure of VClI in Figure \ref{2}a shows the intrinsic semiconducting property with a 2.30 eV band gap which is a little less than 2.72 eV and 2.65 eV for VBrI and VClBr, respectively. In Figure \ref{2}b-e, it is found the indirect gap is formed between valance band maximum (VBM) at $\Gamma$  point and conduction band minimum (CBM) at M point near the Fermi level. $e_{g}$ states of V make a major contribution to the CBM while the VBM is dominated by $p_{x}$ and $p_{y}$ orbits of I (Figure S4 and Table S\uppercase\expandafter{\romannumeral2}).
\subsection{3.3 Competitive Mechanism between FM and AFM.}
\begin{figure}[htbp]
 \centering
 \includegraphics[scale=0.2]{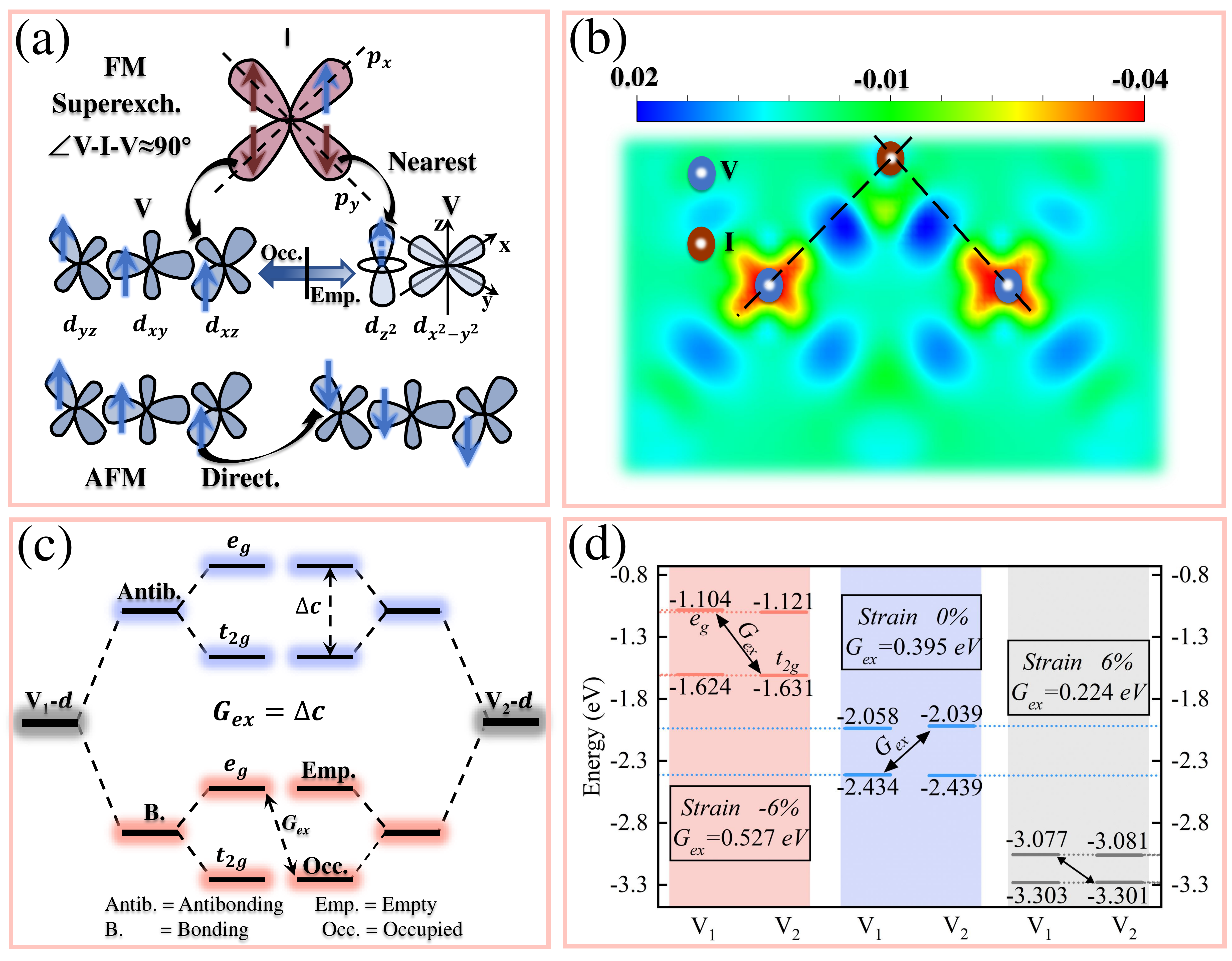}
 \caption{\label{3}(a) Schematic diagram of superexchange and direct exchange for monolayer VClI. (b) The sliced 2D charge density difference for (-1, 2, 1.1) plane. (c) Schematic diagram of exchange interplay and $d$-orbits energy level staggers between V-V atoms. The orange and blue denote bonding and antibonding states, respectively. $\Delta c$ and $G_{ex}$ mean the crystal field splitting and the virtual exchange gap between empty $e_{g}$ and occupied $t_{2g}$ states, respectively. In VXY, theoretically, the value of $G_{ex}$ is equal to $\Delta c$. (d) On-site energies for $d$ orbits between nearest-neighbor V atoms in monolayer VClI under strain -6\%, 0 and 6\%.}
     \end{figure}

Magnetic mechanisms that leading to FM ground state in hexagonal trihalides are complicated. But one can confirm that FM dominates the competition between AFM and FM when the weak second nearest-neighbor interplay is ignored.
On the one hand, two electrons occupying the $t_{2g}$ level take direct interaction with $d-t_{2g}$ states between V atoms, leading to the spin of valence electrons antiparallel configured and the AFM is systematically favored. On the other hand, two nearest-neighbor V atoms take indirect interplay mediated by $90^{\circ}$ angled halogen atoms: one electron locating in V$-d-t_{2g}$ is theoretically excited to the $p_{x}$ or $p_{y}$ orbits of ligands X (Y) (Figure \ref{3}a,b). This superexchange interplay between nearest-neighbor V atoms gives rise to FM spin array in system.\cite{13, 14, 15, 16} Nevertheless, varied superexchange paths exhibit in monolayer crystal containing the nearest-neighbor, 2nd-nearest-neighbor, and 3rd-nearest-neighbor, etc (Figure S12). So it is complicated to take into consideration all the actual coupling effects.

Actually, no matter how many virtual superexchange paths exhibit between V atoms, electrons eventually hop from occupied $t_{2g}$ of one to empty $e_{g}$ states of another. That is, the superexchange tends to be enhanced when the virtual exchange gap $G_{ex}$ between the two energy levels tends to decrease\cite{CHuang2018} (Figure \ref{3}c). To estimate the $G_{ex}$, we use maximally localized Wannier functions (MLWFs)\cite{wannier} to obtain on-site energies for Cr$-d$ and V$-d$ (see Section S\uppercase\expandafter{\romannumeral4} in the supplemental material for details). The $G_{ex}$ of 1.732 eV in the monolayer VI$_{3}$ is less than that in monolayer CrI$_3$ (2.156 eV). More importantly, from the $G_{ex}$ values of 0.395 eV, 0.397 and 0.445 eV for VClI, VBrI and VClI, respectively, it shows the electrons occupying V$_{1}-d-t_{2g}$ in VXY have more change to hybridize with that in empty V$_{2}-d-e_{g}$ states when compared with CrI$_{3}$ and VI$_{3}$. This consequence results in more intensive FM order and higher Curie temperature transforming to paramagnetism: $T_{C(VXY)} > T_{C(VI_{3})} > T_{C(CrI_{3})}$.

\subsection{3.4 Both Weakened Direct and Indirect Exchange Interaction as Stretching}
\begin{figure*}[htbp]
\centering
\includegraphics[scale=0.1]{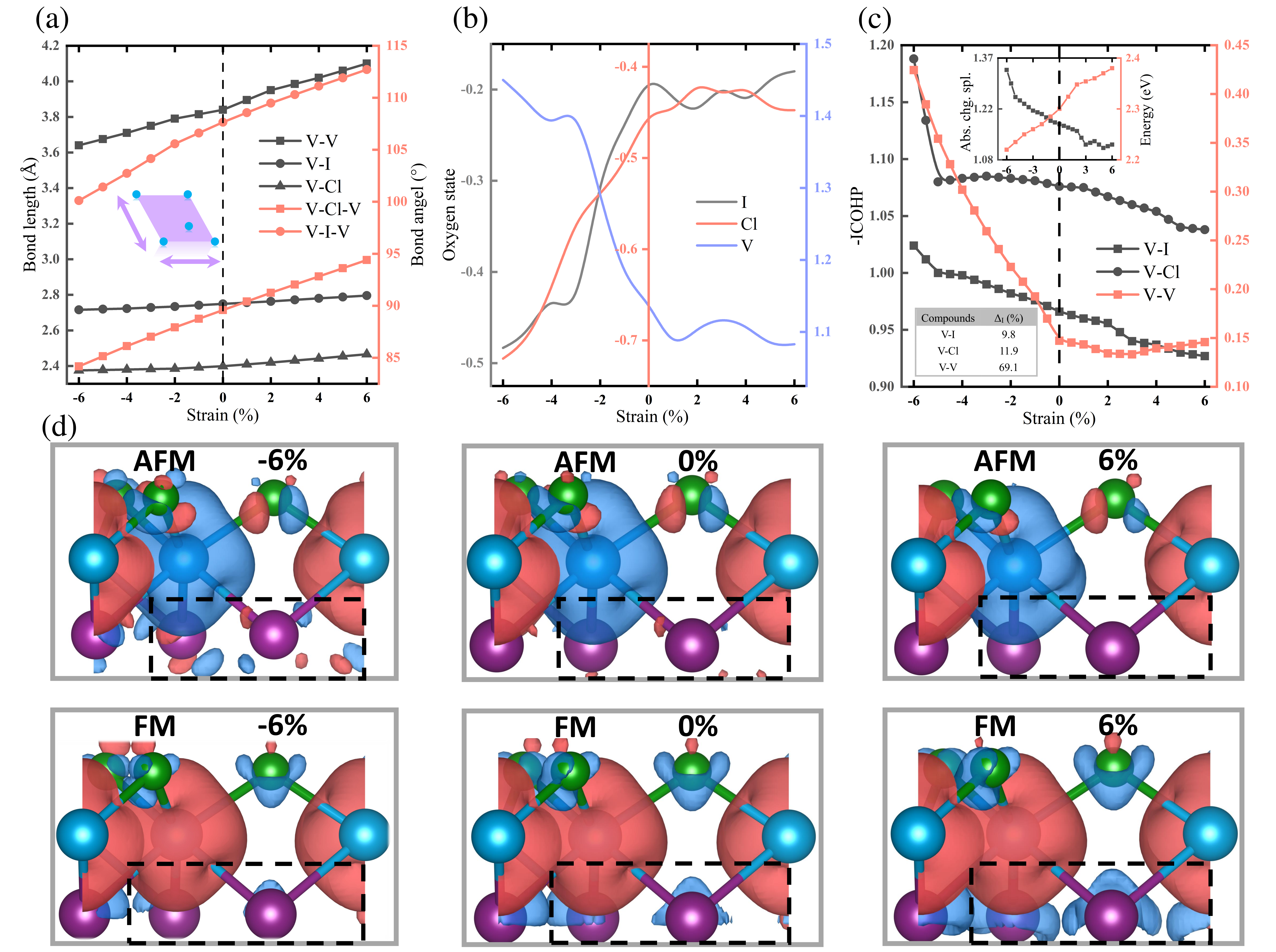}
\caption{\label{4}(a) Bond lengths between nearest-neighbor V-V, V-I, V-Cl and bond angels of V-Cl-V, V-I-V, as a function of strain in monolayer VClI. The inset shows biaxial strain is applied. (b) The average oxygen states of I, Cl and V ions in a unit cell of monolayer VClI. (c) Variation of the average opposite ICOHP and the values are taken between V-Cl, V-I, V-V atoms of monolayer VClI under different strains. The inset shows the absolute average of charge spilling values over occupied bands and the band gaps for monolayer VClI as a function of biaxial strain. $\Delta_{I}$ means the corresponding change rate of -ICHOP.  (d) Spin density for AFM and FM of monolayer VClI. Red and blue denote majority and minority spin channels, respectively. The spin density panels are generated by VESTA package\cite{21} and the isosurface values of all panels are set to 0.0028 eV {\AA}$^{3}$.}
\end{figure*}

Figure \ref{4}a shows the V-V, V-X (Y) bonds are almost proportional to the increasing lattice constants. During the process of enlarging the crystal, the distance between nearest-neighbor V atoms shows faster growth than that between V and X (Y) atoms. The increasing bond lengths indicate the rising exchange barrier between nearest-neighbor V$-d-t_{2g}$ as well as between $t_{2g}$ and $e_{g}$ states. For verification, we apply Bader charge analyses for both the strained and fully relaxed VXY. Figure \ref{4}b plots the tendency for average oxygen state each element in monolayer VClI (see Table S\uppercase\expandafter{\romannumeral3} for Mulliken and L\"{o}ewdin charge analysis). From -6\% to 6\%, the oxygen state for V ion shows a total decreasing trend from 1.45 to 1.08 while I (Cl) keeps rising from -0.72 (-0.48) to -0.44 (-0.22). It implies that the valence electrons of V hopping to I (Cl)-$p$ orbits are reduced by stretching.

In order to quantify the interplay intensity between atoms, we employ crystal orbital Hamilton population analysis (COHP) and the opposite values of ICOHP reflect the intension of exchange effect (Section S\uppercase\expandafter{\romannumeral5} in the supplemental material).\cite{17, 18, 19, 20} As shown in Figure \ref{4}c, it is found that V-I and V-Cl exchange intensities are decreased by 9.8\% and 11.9\%, respectively. This consequence is consistent with the declined number of transition electrons mentioned above. Additionally, the -ICOHP between V atoms is much smaller than V-Cl and V-I, and is drastically declined by 69.1\% from 0.424 to 0.131. The result shows the weakened magnetic coupling effect between nearest-neighbor V atoms (Figure S6).

\subsection{3.5 Enhanced Ferromagnetism in VXY}
A simultaneous descend of direct and indirect interplay is observed. Here a question arises spontaneously, which kind of interplay is reduced more rapidly by tensile strain? From the point of view of electrons hopping, the virtual exchange gap $G_{ex}$ of VXY is found sensitive to strain and is diminished by stretching. In monolayer VClI, the $G_{ex}$ diminishes from 0.527 eV to 0.224 eV under strain from -6\% to 6\% (Figure \ref{3}d, S5). Thus, from the greatly reduced $G_{ex}$, it can be confirmed that the system FM is enhanced by tensile strain. To give further evidence, total energies with different spin arrays are explored. Here six cases of spin configurations are established under biaxial strain and Zigzag array is the most energetically favorable among the AFM configurations (Figure S8, S9). The energy difference between FM and AFM (Figure S10) shows the stable FM ground state without phase transition in the present strain range. Importantly, an overall ascendent tendency in the energy-difference curves of VXY implies the system is increasingly FM favorable from -6\% to 6\%. Besides, from the shapes of spin density isosurfaces (Figure \ref{4}d), the magnetism of VXY is intuitively displayed. $p\sigma$ and $p\pi$ electrons locate originally around I anions in AFM and are absent in the stretching case. Inversely, the unpaired $p$-bonding electrons always appear around I anions in FM configuration. It is clearly shown that the increased system FM and the decreased AFM in monolayer VClI (Figure S7).

We have known the system FM is driven by direct versus indirect exchange interaction between V-V atoms. Along with the consequence that the two coupling effects are simultaneously weakened with tensile strain as discussed above, hence, it can be confirmed that the direct interaction is quenched, but the superexchange one possesses a slower reduction. Which leads to the enhanced FM and the elevated $T_C$ in VXY. This result also yields agreement with the rapidly lengthened V-V bonds and the swiftly declined -ICOHP of V-V atoms under strain from -6\% to 6\% (Figure \ref{4}a,c).

\subsubsection{3.6 Magnetic Anisotropy and Curie Temperature}
\begin{figure*}[htbp]
    \centering
    \includegraphics[scale=0.17]{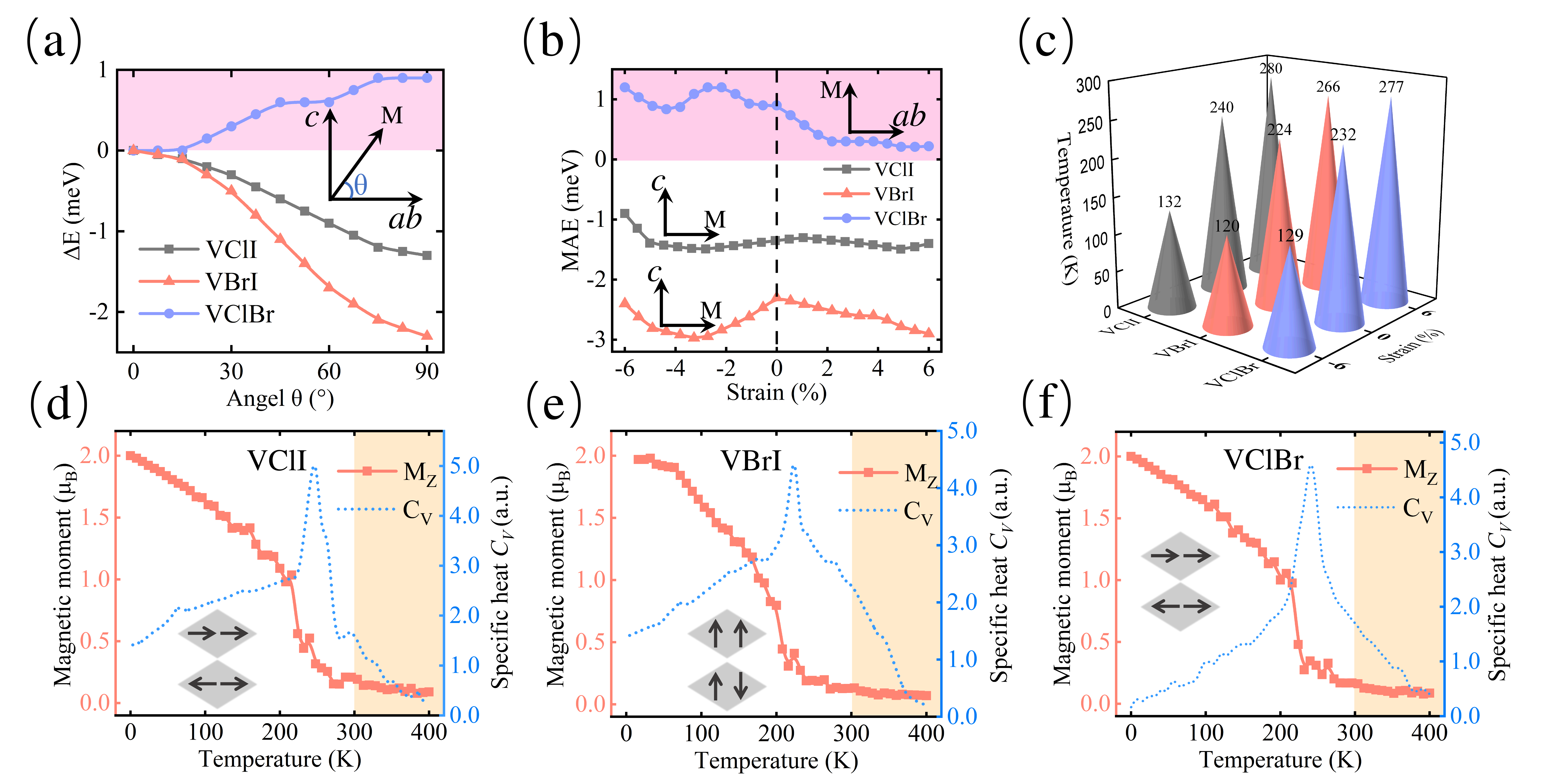}
       \caption{\label{5}(a) Energy difference ($\Delta E$) between $E_{ab}$ and $E_{\theta}$  per unit cell as a function of angel ($\theta$). (b) MAE with respect to biaxial strain. Easy axis for VClBr is perpendicular to $ab$ plane while VClI and VBrI are parallel to $ab$ plane. Purple areas in (a) and (b) denote the magnetic moment direction vertical to $ab$ plane. Additionally, during estimation of the $\Delta E$ and MAE in (a) and (b), spin-orbit coupling effect is taken into consideration. (c) $T_C$ of VXY as a function of strain and the $T_C$ is established based on Monte Carlo simulations with Heisenberg model. (d-f) Magnetic moment per V ion and specific heat ($C_V$) per unit cell as a function of temperature for monolayer (d) VClI, (e) VBrI and (f) VClBr, respectively. Orange denotes the areas above room temperature.}
     \end{figure*}

We now back to estimate the Curie temperature via Monte Carlo simulations of Heisenberg model. We start by confirming the easy axis of VXY that is crucial for studying the magnetic coupling effect between V$-d$ orbits. Figure \ref{5}a plots $\Delta E = E_{ab} - E_{\theta}$ as a function of $\theta$, where $E_{ab}$ and $E_{\theta}$ denote the energies for equilibrium FM state whose spin directions are parallel and $\theta$-angled to $ab$ plane, respectively. Also, 0$^{\circ}$ and 90$^{\circ}$ represent in-plane and out-of-plane spin arrays, respectively. It is found that only VClBr possesses out-of-plane spin array with energetically increasing tendency. While VClI and VBrI favor in-plane array and trend to decrease. Then, VBrI has the largest magnetic anisotropy energy MAE = -2.3 meV (MAE = $E_{ab}$-$E_{c}$, where $E_{c}$ denotes the energy of equilibrium state with out-of-plane spin array) (Figure \ref{5}b). The MAE diagram also shows the stability of easy axis for VXY: VClI possesses a weak coupling effect with strain and the MAE is enhanced by 33\% (from -1.35 eV to -0.9 eV) under compressive strain while declined by 10\% under tensile strain. In VClBr, MAE is asymptotically closed to 0 as the rising tensile strain but reaches convergence from 2.18\%.

In Figure \ref{5}d-f, we plot the results for $T_C$ of equilibrium VXY. The $T_C$ can be extracted as the point where the magnetism vanishes or as the point where the $C_V$ diverges. The two-methods show accordant results for temperature in this case. It is found when compared with the pristine monolayer VI$_3$, the $T_C$ is significantly boosted with values of 240 K, 224 K and 232 K for VClI, VBrI and VClBr, respectively (Section S\uppercase\expandafter{\romannumeral7} in the supplemental material). Although semiconducting property varies slightly with strain (inset of Figure \ref{4}c), $T_C$ possesses a strain-dependent character: the $T_C$ of VXY under 6\% is over twice as much as the values under -6\%. Resulting from the significantly coupling effect with strain, the maximum point (280 K) occurs at VClI under 6\% and is near room temperature (Figure \ref{5}c).

\section{4. Conclusion}
In summary, we reveal a novel general strategy that can elevate $T_C$ for the newly synthesized monolayer VI$_{3}$ to 240 K by replacing the crystal with an asymmetric out-of-plane Janus configuration. It is notable that the $T_C$ of VXY is sensitive to strain and can be further elevated to 280 K by the tensile strain of 6\%. Therefore, combing with the stability analysis, VXY is considered high-temperature and stable intrinsic FM semiconductor. The reasons for the high $T_C$ in VXY and the elevating effect of tensile strain are systematically investigated. The former is due to lower virtual exchange gap $G_{ex}$ than monolayer VI$_{3}$ and the latter originates from direct-interaction rapidly quenching effect with tensile strain. The present work gives one more step forward to enlarge the family of room-temperature FM semiconductors.

\begin{acknowledgement}

This work was supported by National Natural Science Foundation of China (No.11904312 and 11904313), the Project of Hebei Education Department, China (No.ZD2018015 and QN2018012), and the Natural Science Foundation of Hebei Province (No.A2019203507). Thanks to the High Performance Computing Center of Yanshan University.

\end{acknowledgement}

 \begin{figure*}[htbp]
    \centering
    \includegraphics[scale=0.3]{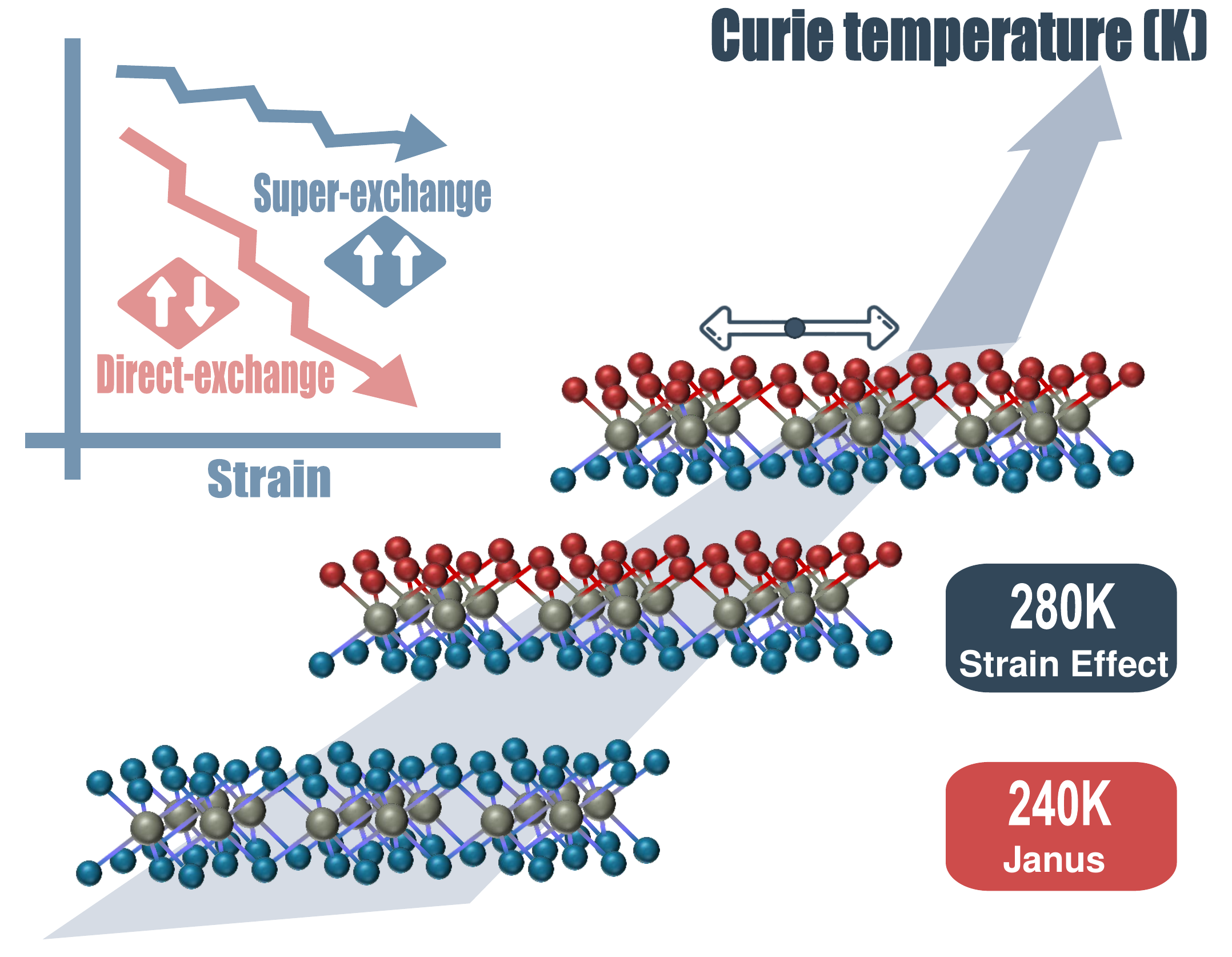}
       \caption{\label{6}For Table of Contents Only}
     \end{figure*}


\end{document}